# First Steps into Physics in the Winery

Roberto Benedetti, Emilio Mariotti and Vera Montalbano

Department of Physics, University of Siena, Italy

Abstract

Physics is introduced as a basic matter in the curricula of professional schools (i.e. schools for agriculture, electronic, electrical or chemistry experts). These students meet physics in the early years of their training and then continue in vocational subjects where many physics' topics can be useful. Rarely, however, this connection between physics and professional matters is quite explicit. Students often feel physics as boring and useless, i.e. very far from their interests. In this kind of schools it is almost always required the physics lab, but it does not always exist. The physics teachers of a local Agricultural Technical Institute asked us to realize a learning path in laboratory dedicated to their students, since in their school the physics lab was missing. This institute is the only public school in the Chianti area specializing in Viticulture and Enology, and attending a further year post diploma, allows the achievement of the qualification of Enologist. We report a learning path realized starting from thermal equilibrium to a full understanding of the measures made with the Malligand's ebulliometer. This device is used for determining the alcoholic strength (alcohol concentration by volume) of an alcoholic beverage and water/alcohol solutions in general. The aim was to make interesting measures of physical quantities, calorimetry and state transitions connecting them to the functioning of an instrument that students use in their professional career. We present our considerations on the students' learning process and on the possibility of extend a similar path. The feedback of students and the interests of their teachers convinced us to go further in this way. We intend in the next future to involve teachers of physics and vocational subjects in the design of a physics curriculum spread over two years in which the main physics topics will be introduced to explain the functioning of tools and equipment used, normally, in the winery.

Physics and technology are closely related. Therefore, in vocational education physics is considered a base for many professional subjects. In Italian professional curricula, physics is planned in early years (2-3 hr/week for one or two years, where many activities are expected in laboratory). Regrettably, current practise is very different from one school to another. Sometimes, laboratory is not properly equipped and usually physics and vocational teachers do not coordinate their educational action. Despite selected physics topics are essential for understanding many professional subjects and practises, the connection almost always remains hidden. In some cases, the relevance of physics to the everyday situation in which the student will ultimately work may not be at all apparent. Thus, students perceive physics as a set of laws very far from their interests, i.e. tedious and useless.

On the other hand, physics teachers often have a professional outlet in vocational schools, especially for those graduates in physics. Thus, they are compelled to work in less favourable conditions with little motivated students.

In the last decades theory and research have been concerned with the relation between motivation and learning (e.g. Alderman, 2008; for a short review focused on physics see Fisher & Horstendahl, 1997). Vocational education enhancement requires to explore and understand how transfer of knowledge can be made more effective in this context (e.g. Guile & Young, 2003; for a discussion on the problem of knowledge in vocational curricula see p. 66).

Moreover, the study of agriculture and related topics can provide a context in which science and mathematics key concepts and skills can be explored in order to enrich students knowledge (Dayley, Conroy, & Shelley-Tolbert, 2001). For these reasons, we developed a learning path in this context for investigate the possibility of improving the motivation of students and teachers and if it can be a way for enhancing learning achievements.

We report a curricular lab designed for the second class (15–16 years) of an Agricultural Technical Institute within the Italian National Plan for Science Degree. These work arose from a request made by physics teachers of the local Agricultural Technical Institute of realizing a learning path in our educational laboratory for their students, since there was no physics lab in their school. The goal was to make interesting measures of physical quantities, calorimetry and state transitions connecting them to the functioning of an instrument that students use in their professional career.

In the next section, we summarize the purpose and the methodological choices of National Plan for Science Degree for explaining the reasons for which this collaboration with the secondary school was made possible and successful. In the following one, we give a description of the school in which we realized the learning path. In order to design an appealing learning path for this kind of students, we found inspiration in the context in which students and teachers usually work. Finally, we describe the learning path in details. Some examples of materials used in class and data from students are reported. In the last section, we discuss the results and give some suggestions for further developments in this school and more in general in vocational education.

## National Plan for Science Degree

In recent decades, it has been detected almost everywhere a consistent decrease of graduates in science disciplines, i.e. Mathematics, Physics and Chemistry.

In order to contrast this trend, Italy launched a large and structured plan funded by the Ministry of Education and Scientific Research named National Plan for Science Degree (Piano Lauree Scientifiche, i.e. PLS) in which more than 30 university in all the country realized actions in order to promote scientific degrees through professional development of teachers and orienting of students essentially by means of laboratory activities (PLS website; for a survey of local PLS activities: for Southern Tuscany see Montalbano, 2012, or for Naples and surroundings see Sassi, Chiefari, Lombardi, & Testa, 2012).

The main strategy and methodologies are the following:

- Orienting to Science Degree by means of training,
- Laboratory as a method not as a place,
- Student must become the main character of learning,
- Joint planning by teachers and university.

The main action is centered on PLS laboratories, designed by university and teachers, in which groups of students (up to 15, better if much fewer) perform experiences in physics laboratory. A characteristic of PLS laboratory is that should not be episodic, i.e. students need to be engaged in this activity for a sufficient time (15 hr or more).

According to the national guidelines, PLS laboratories can have different purposes: Laboratories which approach the discipline and develop vocations, Self-assessment laboratories for improving the standard required by graduate courses, Deepening laboratory for motivated and talented students (e. g. Benedetti, Mariotti, Montalbano, & Porri, 2011; Di Renzone, Frati, & Montalbano, 2011).

The activity described in the following can be classified as a laboratory which approach physics and its purpose is to increase the scientific literacy in students who usually have little interest in physics.

**A Peculiar School**

The Agricultural Technical Institute Ricasoli is one of 10 Italian technical special secondary schools specialized in Viticulture and Enology and the only one in Tuscany. After a course of five years, students become expert in Viticulture and Enology and, with a further annual course, are qualified as winemaker technician (Ricasoli Website).

The school was born in 1952 and occupies an area of 47 hectares (116 acres) and deals with managing the educational farm. The Villa of Partini (Sienese architect 1842-1895) hosts the School's Management, the offices of the Secretariat and the Library of the School from which you can dominate the whole area of Chianti and the city of Siena.

Each last class is responsible for a vineyard and students treat all aspects from the cultivation of the vine to the grape harvest, from fermentation to bottling, oversee and determine what is needed to ensure the quality of the final product.

Every year, the last day of school is devoted to the Enological Day. Parents, students, everyone is invited to taste the products made by students during the year. Wine tasting ends with the awarding of the winner of the best wine produced from the last classes.

Therefore, students live in a context where wine culture is deeply embedded with countryside, local history and economy, social tradition and likely their future profession.

For these reasons, we proposed a learning path centered on the use of a professional device routinely utilized by any enologist: The Malligand ebulliometer or ebullioscope (Malligand, 1876).

### From Thermal Equilibrium to Alcoholic Strength Measurement

In the following we give elements useful for understanding how we designed the learning path in the way we did. We describe methods, materials, the organization of activities and how we thought to evaluate the effectiveness of this learning path and participants. The next section would be dedicated to analyze data and findings.

**Physics Contents**

**The alcoholic strength measurement.** Wine and each alcoholic beverages obtained by distillation can be regarded as pure water/alcohol solutions from the physical point of view. All the other components, although important from the organoleptic point of view, are present in quantities which are too small for having an influence on the measurement of alcoholic strength. The alcoholic strength can be determined for ebullioscopy, valid for commercial labelling and trade and which uses a special device: Malligand ebulliometer.

The method is based on different temperature of boiling water (100 °C) and ethanol (78.3 °C), which together give a mixture which has a boiling point intermediate between that of the two substances. Wine is a hydro-alcoholic mixture and has a boiling point which decreases with the increase in content in ethanol.

**The Malligand ebulliometer**. The apparatus is showed in Figure 1 and the labeling in the text refers to the schematic drawing on the right. The Malligand ebulliometer comprises a metal boiler $F$ connected beneath to an annular tube which is inclined and welded to a small chimney $S$ beneath which a heating lamp $L$ is positioned. This annular tube makes it possible for the liquid present in the boiler to be heated by thermal siphoning. The boiler is closed with a screwed lid provided with holes and has a metal arm $E$ bent into a right angle. A thermometer $T$ passes through the central hole in the lid and its bulb dips into the boiler while its capillary, which is also bent into a right angle, is housed horizontally in a metal arm, against a graduated scale $c$. A cooling unit $B$, whose function is to cause condensation of the alcoholic vapours to prevent any change in concentration of the solution altering the boiling point, is housed in the side hole of the cover.

**Learning Path Description.**

**Overview.** The measurement of alcoholic strength needs elements of thermal exchanges, properties of matter in function of temperature and changes of state properties. Thus, we decided to design a learning path that started by characterizing thermal equilibrium, gave some elements of calorimetry, analyzed temperature in a system where a constant amount of heat is supplied and finally utilized the Malligand ebullioscope. Despite PLS indications, we had no enough time for joint planning with teachers. They contributed to the timing optimization of labs and lessons, discussed with their students in classroom the concepts introduced and the data obtained but did not participate to initial educational choices.

**Methods.** The learning path was designed as a PLS laboratory realized during school hours. We planned 3 activities in laboratory (each one of 3 hr for a total of 9 hr) and a final participated lesson for summarizing the concepts encountered in the labs also through data analysis, connecting them with their current knowledge and their future professional practice (3 hr). The first lab was performed in Department of Physics, while all other activities were realized at the school (microbiology laboratory and classroom). In the laboratory, students were assisted by one or two of the authors (V. M. and R. B.) with the collaboration of the teachers who were accompanying the class (usually 2 or 3, i.e. a physics teacher, a technical assistant and/or a support teacher for students with special needs). The final lesson was held by the author with more experience in winemaking and winery (R. B.) assisted by another author (V. M.) while teachers attended. We designed some materials, described in the next section, for facilitating comprehension of students, supporting them in laboratory activities and in preparing of final reports. We planned to evaluate the effectiveness of this action by means of direct observations performed by the authors during the activities with students, interviews with teachers for understanding the impact on motivations and learning achievements and, finally, by analysing students final reports.

**Materials.**

*Laboratory on thermal equilibrium.* Students worked in small groups, but individual reports were requested. They started by observing the thermal equilibrium between solids (same mass and material, different mass and same material, same mass and different material) and solid/water in a Dewar flask. Then, a measure of heat capacity was realized. We focused on measures and error evaluations. Since this first experience in laboratory was very intense, we prepared a detailed worksheet in which all activities were explained, with some hints (e. g. there was indicated where to write a measure with spaces for error and measure units, space for calculation, data analysis and discussion). Therefore, students could follow the activity and, by completing the worksheet in class and at home, they had a ready report for assessment.

*Laboratory on changes of state.* The activity was to study the behaviour of water and water-alcohol solution at the boiling point and to compare them. We did not give any worksheet in this case.

*Laboratory on the Malligand ebulliometer.* Since the boiling temperatures depend on the local atmospheric pressure, the zero point of scale must be fixed. Water was placed in the boiler of Malligand's device, the lamp was turned on and the zero of the sliding scale was fixed when it was coincident with the point of maximum extension reached by the mercury's meniscus in the capillary. Then, the water was replaced with the same water/alcohol solution that they had examined in the previous lab. The device was heated again, and the value in alcoholic degrees could be read directly on the scale, in correspondence with the maximum point reached by the mercury.

Also in this case, we did not give any worksheet and we required a final report with data analysis and discussion at the end of the learning path (i. e., after the final lesson).

*Final lesson.* After all the activities in laboratory, students were guided in understanding their measures and physics processes involved in Malligand ebulliometer by a guided discussion and by a set of questions such as:

- By analyzing the time-temperature graph relative to boiling water can you deduce if and when the temperature is proportional to the energy supplied by the heater?

- What can be inferred in the case of boiling mixture, comparing to what happened in the case of water alone?

- The tool that you know more similar to Malligard's device is ...................... Explain why.

- Can you explain why the scale is not linear?

- You have a graph which shows the boiling temperature of a mixture of water-alcohol as a function of composition of the mixture; how it is connected to the non-linearity of the scale of the Malligand's device?

- The alcohol content measured with the Malligand ebullioscope is greater when the meniscus has less displacement. Do not you think it works differently than other scales?

All students received the complete set of questions and the chart of boiling point in function of composition of liquid mixture (similar to the one shown in Figure 2) in order to get some hints on all important aspect in the use of Malligand ebullioscope that can be inserted properly in the final report.

Our intent was that the two individual reports were collected by teachers for assessment of students at school. Moreover, we asked to have back all reports at the end of scholastic year for evaluating the effectiveness of the learning path.

**Participants.** All second classes of the Agricultural Technical Institute participated (5 classes, about 100 students aged 15-16 years).

**Data and findings.** Since the learning process lasted until near the end of the scholastic year, it was not possible to obtain the final reports of the students (only 20 reports of the first laboratory and 13 final reports of a single class were available, while all the others were missing although they were assessed by teachers). Thus, it was impossible to make directly a quantitative assessment of learning findings and we must limit ourselves to what emerged from direct observations through the activity, interviews with teachers and few observations on the available reports. In the laboratory as well as in conclusive lesson in class, students were interested and active, even though the level of attention could fall during complex tasks, especially theoretical one. Most of them were able to put data in a graph and connecting it to phenomena (seen and unseen directly). In Figure 3, data from a group of students is showed. The graph refers to the second experience in laboratory and during discussion in class many students were able to connect properly the time of heating of a constant heater, heat transferred to the liquid, variations of temperature, changes of state and how Malligand's device works. Students were usually very careful in performing activities in laboratory, as shown in Figure 4, where a group was following the mercury's meniscus in the capillary by using a flashlight as suggested by some of them. During the discussion, few students seemed to have captured the relationship between data on boiling point of a mixture, showed in Figure 2, and the non-linearity in the scale of the Malligand ebullioscope. However, more students were able to understand how the atmospheric pressure affects the measurement of the alcoholic strength and the importance of the initial set of zero point. Teachers reported a wide interests of pupils, especially for the part in laboratory, and gave a judgment generally very positive for the activity.

## Discussion and Conclusions

The main purpose of increasing students' attention on physics has been fully achieved. As soon as students realized the relationship with some aspects of enology, their involvement increased significantly. A first analysis of students' reports on laboratory experience shows the necessity of paying more attention for integrating this activity in the previous knowledge of students. Another aspect that negatively affected the learning process was the lack of involvement of teachers in the initial design. This fact has led to a lack of involvement in making decisions during the educational process and a little incisive action towards students.

Anyway, these results convinced us to expand the topics of physics that can lead to activities directly usable in technical matters. Actually, we are designing and realizing learning paths, such as mechanics in the winery (winepress), fluid mechanics (hydraulic press, decanting wine, vats' usage), optical (grape refractometer), in close collaboration with physics and vocational teachers in the same school. The idea is of preparing and testing learning paths spread over two years in which the main physics topics would be introduced to explain the functioning of tools and equipment used, normally, in the winery.

## Acknowledgement

This work is based on activities and experiences which were realized within the National Plan for Science Degree supported by Italian Ministry of Education, University and Research. The authors would like to thank the technical staff of Department of Physics for the availability and continuous support always essential for maintaining fully operational physics laboratories.

References

Alderman, M. K., (2008), *Motivation for achievement: Possibilities for teaching and learning*, 2nd ed., New York: Routledge.

Benedetti, R., Mariotti, E., Montalbano, V., & Porri A., (2011), Active and cooperative learning paths in the Pigelleto's Summer School of Physics, *Twelfth International Symposium Frontiers of Fundamental Physics (FFP12)*, Udine 21-23 November 2011, arXiv:1201.1333v1 [physics.ed-ph]

Dayley, A. L., Conroy, C. A., & Shelley-Tolbert, C. A., (2001), Using agricultural education as a context to teach life skills, *Journal of Agricultural Education*, *42*(1), 11-40.

Di Renzone, S., Frati, S., & Montalbano, V., (2011), Disciplinary knots and learning problems in waves physics, *Twelfth International Symposium Frontiers of Fundamental Physics (FFP12)*, Udine 21-23 November 2011, arXiv:1201.3008v1 [physics.ed-ph]

Fisher, H. E., & Horstendahl, M., (1997), Motivation and learning physics, *Research in Science Education, 27*(3), 411-424.

Guile, D., & Young, M., (2003), Transfer and transition in vocational education: Some theoretical considerations, in T. Tuomi-Gröhn & Y. Engertröm (Eds.), *Between school and work. New perspectives on transfer and boundary-crossing* (pp. 63-81), Oxford. UK: Elsevier Science.

Malligand, P. M. B., (1876), Improvement in ebullioscopes, Patent number 173128, http://www.google.com/patents/US173128, accessed 2012 November

Montalbano, V., (2012), Fostering Student Enrollment in Basic Sciences: the Case of Southern Tuscany, in *Proceedings of The 3rd International Multi-Conference on Complexity, Informatics and Cybernetics*: IMCIC 2012, ed. N. Callaos et al, 279.

PLS website http://www.progettolaureescientifiche.eu/, accessed 2012 June

Ricasoli Website, www.istitutoagrario.siena.it/index.htm, accessed 2012 June.

Sassi, E., Chiefari, G., Lombardi, S., & Testa, I. (2012), Improving scientific knowledge and laboratory skills of secondary school students: the Italian Plan "Scientific Degrees",

Poster Session Strand 3: Learning Physics Concepts, P2.G02.03, WCPE

WIKI Chart, Vapour-liquid equilibrium mixture of ethanol and water, http://en.wikipedia.org/wiki/File: Vapor-Liquid_Equilibrium_Mixture_of_Ethanol_and_Water.png, accessed November 2012

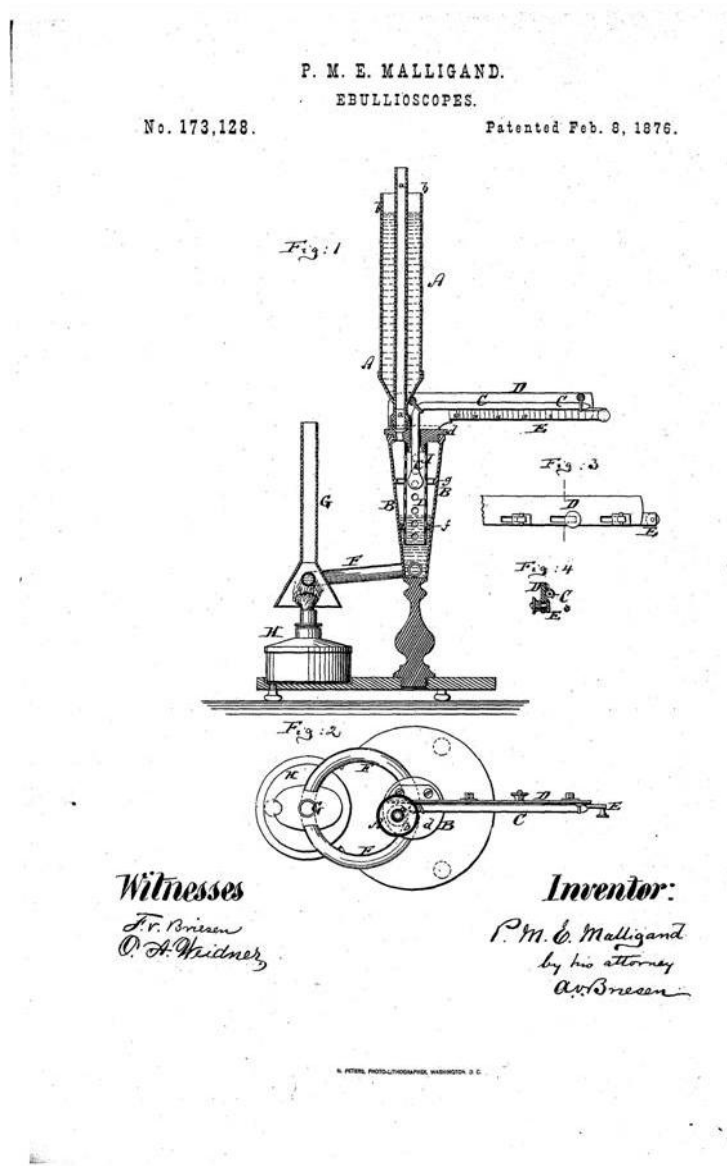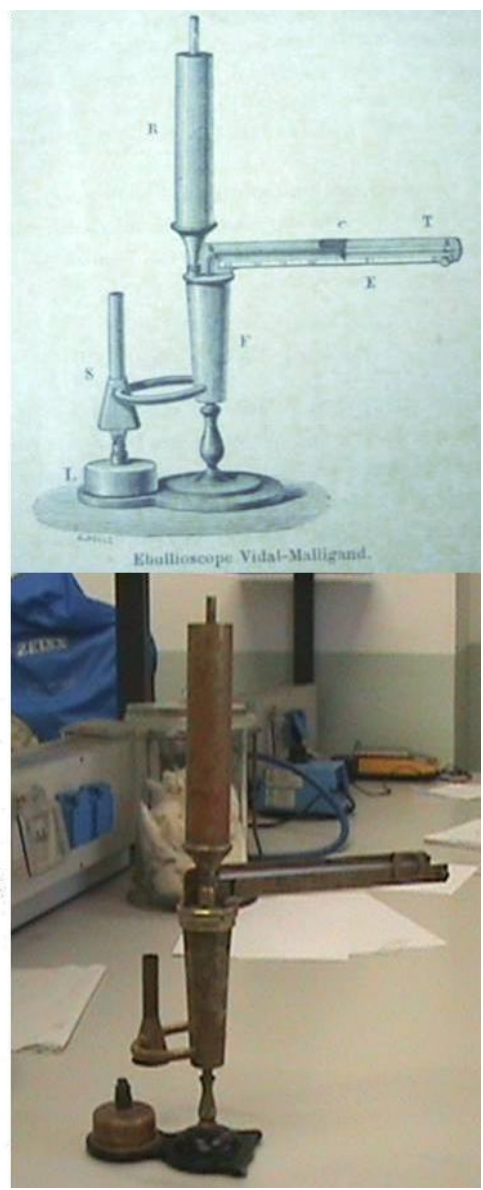

*Figure 1.* On the left, the original drawing of the Malligand ebullioscope in the patent is shown. On the right, a schematic drawing, with the labelled parts cited in the text, is shown and a picture of the devise utilised by students is given.

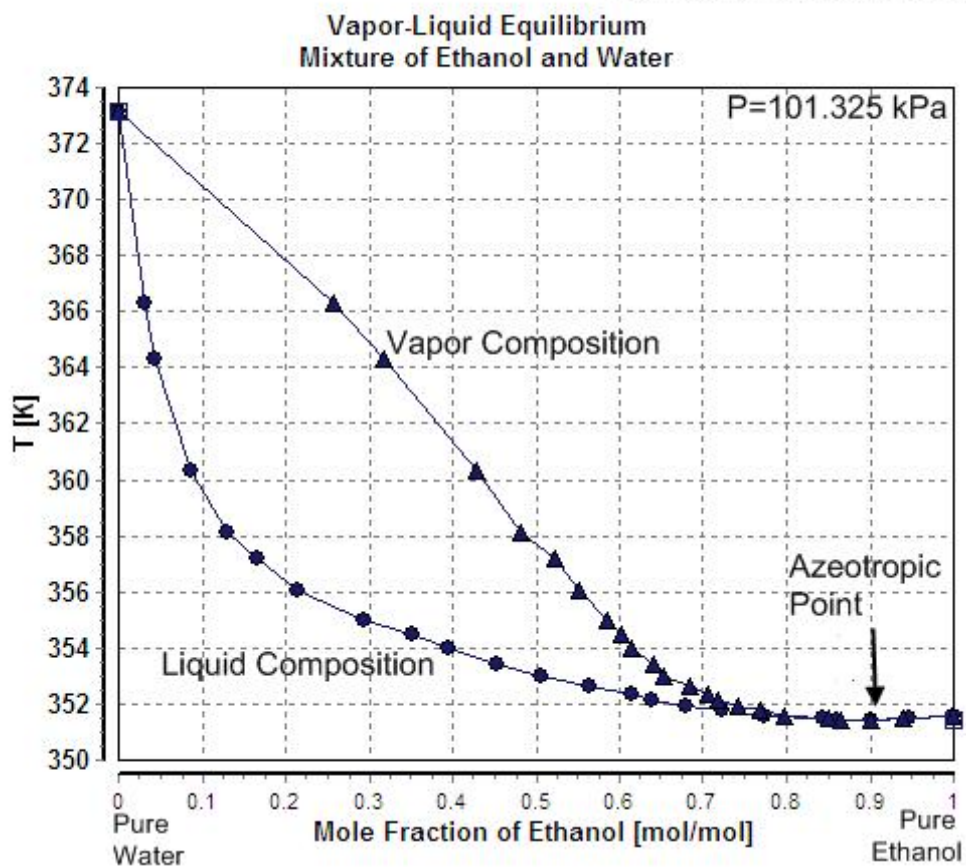

*Figure 2.* Boiling point in function of liquid composition of a mixture of ethanol and water at a fixed pressure (WIKI Chart).

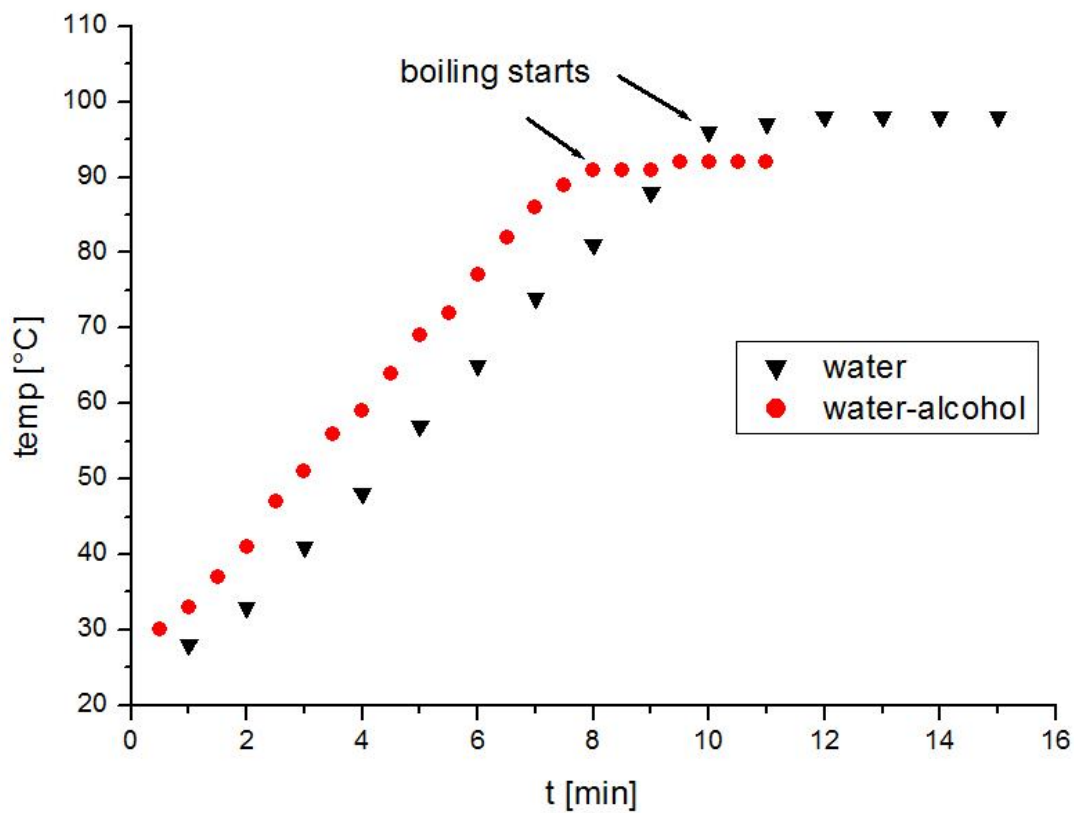

*Figure 3.* Temperature dependence from time is showed, when constant heat is supplied to the liquid. The data were collected by a group of students during the laboratory on changes of state.

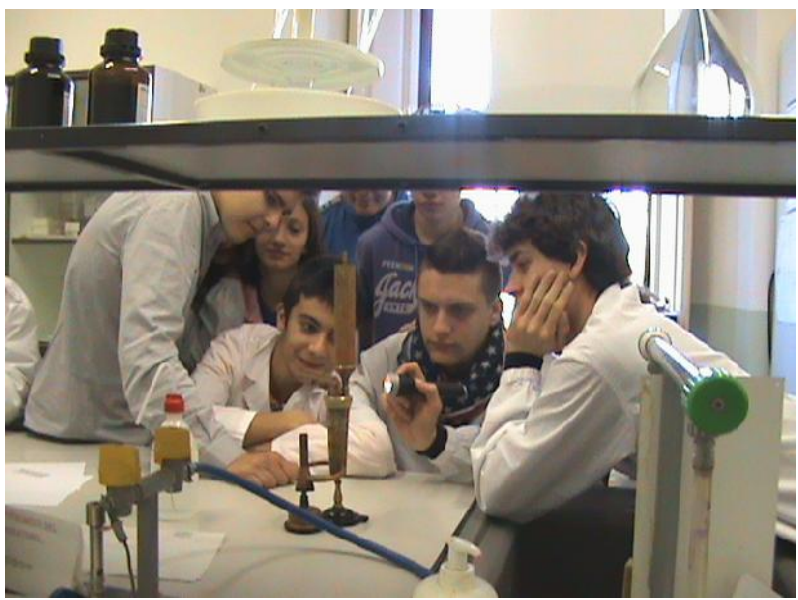

*Figure 4.* Students are carefully following the mercury's meniscus on the Malligand's device aided by a flashlight.